\documentclass[]{aa}
\usepackage{graphicx}
\usepackage{natbib}
\usepackage{ulem}
\bibpunct{(}{)}{;}{a}{}{,}
\bibpunct{(}{)}{;}{a}{}{,}
\usepackage{txfonts}
\begin{document}
   \title{APEX 1\,mm line survey of the Orion Bar}

   \author{S. Leurini\inst{1}, R. Rolffs\inst{1},  S. Thorwirth\inst{1}, B. Parise\inst{1}, 
P. Schilke\inst{1}, C. Comito\inst{1}, F. Wyrowski\inst{1},  R. G\"usten\inst{1}, P. Bergman\inst{2}, K. M Menten\inst{1}, \and
          L.-\AA. Nyman \inst{2}
          }

   \offprints{S. Leurini}

   \institute{Max-Planck-Institut f\"ur Radioastronomie,
              Auf dem H\"ugel 69, 53121 Bonn, Germany\\
              \email{sleurini@mpifr-bonn.mpg.de}
         \and
             European Southern Observatory, Casilla 19001, Santiago, Chile
             }

   \date{}
 \abstract
   {Unbiased molecular line surveys are a powerful tool for analyzing  the physical and chemical parameters of astronomical objects and are  the only means for obtaining a complete view of the molecular
inventory for a given source. The present work stands for the  first such investigation of
a photon-dominated region.
   }
   {The first results of an ongoing millimeter-wave survey obtained towards the  Orion Bar are reported.
   }
   {The APEX telescope in combination with the APEX-2A facility  receiver was employed in this investigation.
   }
   {We derived the physical parameters of the gas through LVG analyses of  the 
methanol and formaldehyde data. Information on the sulfur and  deuterium chemistry of photon-dominated regions is obtained from  detections of several sulfur-bearing molecules and DCN.
   }
   {} 
   \keywords{ISM: individual objects (Orion Bar) - ISM: abundances - ISM: molecules}

\authorrunning{Leurini et al.}

   \maketitle
%

\section{Introduction}

Photon-dominated regions (PDRs)  are commonly found interstellar
environments whose global properties are determined by intense
far-ultraviolet (FUV) radiation emerging from nearby young OB stars
\citep[e.g.][]{hollenbach_ARA&A_35_179_1997}.  Owing to its proximity
and nearly edge-on orientation, the Orion Bar PDR has received
particular attention as a template for studies of the spatial
stratification of various atomic and molecular species from the highly
penetrated surface layers deep into the parental molecular cloud.
Targeted (sub)millimeter-wave investigations indicate
that, besides a considerable number of ubiquitous astronomical
molecules such as CO, CS, HCN, CH$_3$OH, and H$_2$CO, other species can
be found that are suggestive of a unique PDR chemistry
\citep{hogerheijde_AaA_294_792_1995,jansen_AaA_303_541_1995,fuente_AaA_406_899_2003}.
One such example is the molecular ion CO$^+$
\citep[e.g.][]{stoerzer_A&A_296_9_1995}, and another one appears to be
CF$^+$, which so far has only been detected toward the Orion Bar
(Neufeld et al 2006, this volume).  PDRs are also thought to
show enhanced abundances of molecules carrying refractory elements due
to grain breakup caused by the FUV radiation field
\citep{schilke_A&A_372_291_2001}.  Additionally, observations of the
Horsehead nebula suggest that at least part of the molecular carbon
chain budget in PDRs may be produced from UV-destruction of PAHs and
carbonaceous grains \citep{pety_AaA_435_885_2005}.

In view of this, PDRs are attractive targets for
unbiased molecular line studies, which also help to derive the global picture of the
physics and chemistry associated with them.  Here, we present the initial
results from an unbiased molecular line survey obtained between 279
and 308\,GHz with the APEX telescope (G\"usten et al. 2006, this volume).

\section{Observation}

Observations were conducted with the APEX~12-m telescope\footnote{This
   publication is based on data 
   acquired with the Atacama Pathfinder Experiment 
   (APEX). APEX is a collaboration between the Max-Planck-Institut f\"ur Radioastronomie, 
   the European Southern Observatory, and the Onsala Space Observatory.} towards the Orion Bar
at the position $\alpha_{2000}=5^h35^m25.3^s$,
$\delta_{2000}=-5^\circ24^\prime 34.0^{\prime\prime}$, corresponding to the
``Orion Bar (HCN)'' position of \citet{schilke_A&A_372_291_2001}, the most massive clump seen in H$^{13}$CN
\citep[][ hereafter LS03]{lis_ApJL_597_L145_2003}.
Data were taken in 2005,
between October and December in the position-switching mode, with the reference position at ($600''$, 0$''$). 
The APEX-2A 
receiver (Risacher et al. 2006, this volume) was used in combination with
the Fast Fourier Transform Spectrometer (Klein et al. 2006, this volume)
providing a bandwidth of 1\,GHz and a resolution of 0.12~km~s$^{-1}$.
The frequency range between 279 and 307.7~GHz was completely covered; additional 
data were taken in selected frequency ranges from 318.5~GHz to 361.5~GHz. 
Step-widths of approximately 500\,MHz were
used to cover each frequency setup twice to facilitate the sideband assignment. 
The  antenna temperature was converted to the main-beam brightness temperature
by using forward and beam efficiencies 
of 0.97 and 0.74, respectively. The noise level is of the order  of 0.06~K--0.1~K, and the
system temperature around 160~K.
The calibration was performed by using the APECS software (Muders et al. 2006, this volume).
The  pointing was checked on Mars and was found to be accurate within a few arc-seconds.
We estimate a calibration uncertainty of $\sim 30\%$. 

\section{Analysis of the data}
The line identification was based on the Cologne Database for
Molecular Spectroscopy\footnote{http://www.cdms.de} \citep{mueller_cdms} and on the
JPL\footnote{http://spec.jpl.nasa.gov} \citep{pickett_JMolSpectrosc_60_883_1998}
line catalog and performed with the XCLASS software. We identified
16 different species and a number of their isotopologs; no strong unidentified lines were found 
in the range surveyed.
In the following paragraphs, we focus on the excitation conditions of methanol (CH$_3$OH)
and formaldehyde (H$_2$CO), on the one hand, and  on the analysis of S-bearing molecules, on the other.
Methanol and formaldehyde are present in our dataset with a large number of transitions (Table~\ref{lineid}), and their collisional rates are available \citep{1991ApJS...76..979G,2002JPhB...35.2541P}. Thus, a more 
 rigorous multi-line analysis is possible for these two species, which delivers temperature and density estimates. 
As for the S-bearing molecules in our dataset, some of them, namely CS, H$_2$S, SO, SO$^+$,  and SO$_2$, have already  been
studied towards this source \citep{hogerheijde_AaA_294_792_1995,jansen_AaA_303_541_1995,fuente_AaA_406_899_2003},
whereas HCS$^+$, H$_2$CS and NS are being analyzed here for the first time in this environment. A comprehensive analysis will be presented once the line survey has been completed.

\begin{table}[h!]
\caption{Line parameters of the observed transitions}\label{lineid}
\begin{tabular}{lcrr}
\hline\hline
\multicolumn{1}{c}{transition}&\multicolumn{1}{c}{frequency} &\multicolumn{1}{c}{E$_{low}$}&\multicolumn{1}{c}{$\int T_{MB}dv$}\\
&\multicolumn{1}{c}{[MHz]}&\multicolumn{1}{c}{[K]}&\multicolumn{1}{c}{[K km s$^{-1}$]}\\

 H$_2$CO        $  4_{1,4}   \to      3_{1,3}      $    &        281526.929 &  32.06&$10.7 \pm 0.5$\\
 H$_2$CO        $  4_{0,4}   \to      3_{0,3}      $    &        290623.405 &  20.96&$5.9\pm 0.4$\\
 H$_2$CO        $  4_{2,3}   \to      3_{2,2}      $    &        291237.767 &  68.09&$2.4 \pm 0.3$\\
 H$_2$CO        $  4_{3,2}   \to      3_{3,1}      $    &        291380.488 & 126.95&$3.4\pm 0.3$\\
 H$_2$CO        $  4_{3,1}   \to      3_{3,0}      $    &        291384.264 & 126.95&$3.7\pm 0.3$\\
 H$_2$CO        $  4_{1,3}   \to      3_{1,2}      $    &        300836.635 &  33.45&$9.8 \pm 0.5$\\
 H$_2$CO        $  5_{1,5}   \to      4_{1,4}      $    &        351768.645 &  45.57&$10.1\pm 0.9$\\

 CH$_3$OH-$E$   $  6_0       \to      5_0          $    &        289939.477 &  47.93&$0.5 \pm 0.1$\\
 CH$_3$OH-$E$   $  6_{-1}    \to      5_{-1}       $    &        290069.824 &  40.39&$0.9 \pm 0.2$\\
 CH$_3$OH-$A$   $  6_0       \to      5_0          $    &        290110.666 &  34.82&$1.2 \pm 0.2$\\
 CH$_3$OH-$E$   $  6_{1}     \to      5_{1}        $    &        290248.762  & 55.87&$0.3 \pm 0.2$ \\
 CH$_3$OH-$E$   $  6_2       \to      5_2          $    &        290307.376 &  60.72&$0.3^a \pm 0.2$ \\
 CH$_3$OH-$E$   $  6_{-2}     \to      5_{-2}          $    &    290307.643 &  57.07&\\
 CH$_3$OH-$A$   $  1_1       \to      1_0          $    &        303366.890 &   2.32&$0.6 \pm 0.2$\\
 CH$_3$OH-$A$   $  2_1       \to      2_0          $    &        304208.350 &   6.96&$0.8 \pm 0.2$\\
 CH$_3$OH-$A$   $  3_1       \to      3_0          $    &        305473.520 &  13.93&$0.9 \pm 0.2$\\
 CH$_3$OH-$A$   $  4_1       \to      4_0          $    &        307165.940 &  23.21&$1.4 \pm 0.3$\\
C$^{34}$S   $  6             \to      5          $         &         289209.068 &  34.70  &$  4.6 \pm 0.7$\\
C$^{33}$S   $  6             \to      5          $         &         291485.935 &  34.98  &$  1.6 \pm 0.3$\\
CS          $  6             \to      5          $         &         293912.087 &  35.27  &$ 40.7 \pm 2.2$\\
CS          $  7             \to      6          $         &         342882.850 &  49.37  &$ 33.1 \pm 1.9 $\\

HCS$^+$     $  7             \to      6          $         &         298690.453 &  43.01  &$  0.3 \pm 0.1$\\
HCS$^+$     $  8             \to      7          $         &         341350.229 &  57.34  &$  0.6 \pm 0.1$\\

SO          $  7_6           \to      6_5        $         &         296550.064 &  50.66  &$  3.7 \pm 0.5$\\
SO          $  7_7           \to      6_6        $         &         301286.124 &  56.50  &$  4.0 \pm 0.6$\\
SO          $  7_8           \to      6_7        $         &         304077.844 &  47.55  &$  7.2 \pm 1.3$\\
SO          $  8_7           \to      7_6        $         &         340714.155 &  64.89  &$  3.6 \pm 0.7$\\
SO          $  8_8           \to      7_7        $         &         344310.612 &  70.96  &$  3.0 \pm 1.1$\\
SO          $  8_9           \to      7_8        $         &         346528.481 &  62.14  &$  7.2 \pm 1.2$\\

SO$^+$ $^2\Pi_{1/2}$  $  13/2         \to      11/2e     $ &         301361.501 &  38.92  &$  0.6 \pm 0.1$\\
SO$^+$ $^2\Pi_{1/2}$  $  13/2         \to      11/2f     $ &         301736.791 &  39.03  &$  0.4 \pm 0.1$\\

H$_2$CS        $  9_{1,9}         \to      8_{1,8}      $  &         304307.645 &  71.60  &$  0.6 \pm 0.1$\\
H$_2$CS        $ 10_{0,1,0}       \to      9_{0,9}      $  &         342946.335 &  74.13  &$  0.4 \pm 0.1$\\
H$_2$CS        $ 10_{2,9}         \to      9_{2,8}      $  &         343322.111 & 126.83  &$  0.4 \pm 0.1$\\
H$_2$CS        $ 10_{1,9}         \to      9_{1,8}      $  &         348534.225 &  88.47  &$  0.8 \pm 0.2$\\

NS $^2\Pi_{1/2}$$  13/2      \to      11/2e    $           &         299700.097 &  39.82  &$  0.4 \pm 0.1$\\
NS $^2\Pi_{1/2}$$  13/2      \to      11/2f    $           &         300098.057 &  39.93  &$  0.4 \pm$ 0.1\\
NS $^2\Pi_{1/2}$$  15/2      \to      13/2e    $           &         345823.532 &  54.20  &$  0.7 \pm 0.2$\\
NS $^2\Pi_{1/2}$$  15/2      \to      13/2f    $           &         346220.774 &  54.33  &$  0.8 \pm 0.2$\\

H$_2$S         $  3_{3,0}         \to      3_{2,1}      $  &         300505.560 & 154.48  &$  0.5 \pm 0.1$\\
\hline\hline

\end{tabular}
\begin{list}{}{}
\item $^a$ blended with the $6_{-2}\to 5_{-2}$  CH$_3$OH-$E$ line.
\end{list}
\end{table}

\subsection{CH$_3$OH and H$_2$CO}
The excitation analysis of CH$_3$OH and H$_2$CO was performed by
using the method described by \citet{leurini_A&A_422_573_2004}, 
recently modified to treat H$_2$CO as well. The procedure is based on the simultaneous
fit of all the lines with a synthetic spectrum computed in the LVG
approximation with the cosmic background as the only external radiation
field.  If lines are optically thick, the model fits source size, temperature, H$_2$ density, and CH$_3$OH/H$_2$CO column 
density; for optically thin lines,  source size and column density cannot be determined independently, and beam-averaged column densities are provided. 

To improve
the signal-to-noise ratio in the weaker 
transitions, we smoothed the 
data to a resolution of
0.5~km~s$^{-1}$. At this resolution, each line has a single-peak
Gaussian profile ($v_{\rm{lsr}}=10.0 \pm 0.2$~km~s$^{-1}$, $\Delta v=1.7 \pm 0.3$~km~s$^{-1}$), while, at the original resolution of 0.12~km~s$^{-1}$,
two peaks were detected in the $ 5_{1,5}\to4_{1,4}$~H$_2$CO and in the $
6_{-1}\to 5_{-1}$~CH$_3$OH-$E$ transitions. Non-Gaussian profiles were
found in other transitions as well.  By fitting the $ 6_{-1}\to
5_{-1}$~CH$_3$OH-$E$ line with the GAUSS method of CLASS, we found two
velocity components at $v_{\rm{lsr}}=10.89 \pm 0.06$ and
$v_{\rm{lsr}}=9.82 \pm 0.05$~km~s$^{-1}$, with $\Delta v=0.62 \pm 0.13$ and $\Delta
v=0.63 \pm 0.06$~km~s$^{-1}$. These values resemble what
LS03 found for the H$^{13}$CN clumps 
1, 5 and 8, 
which  are all
within our beam.
For the sake of simplicity, one velocity component
 was used
both for the CH$_3$OH and  the H$_2$CO analysis, as derived from the smoothed CH$_3$OH data. 

\begin{figure}
\centering
\includegraphics[width=5cm,angle=-90,bb=160 10 550 688,clip]{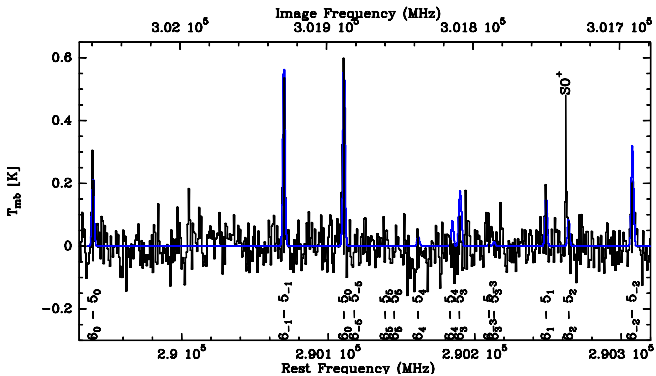}
\caption{Spectrum of the $6_k\to 5_k$ CH$_3$OH band. 
Overlaid on the data, in blue, is the synthetic spectrum corresponding to our best fit.}\label{ch3oh-band}
\end{figure}

High density $(n\sim7\times 10^6$~cm$^{-3}$) and  a moderately high
temperature $(T\sim 70$~K) are needed to reproduce the CH$_3$OH
spectrum. The temperature is constrained to this value by the non-detection of high-excitation (E$_{low}\ge80$~K) 
lines (Fig.~\ref{ch3oh-band}).
The high density value we derived  suggests that CH$_3$OH
comes from small clumps. 
Therefore, guided by
LS03, we used a source size of $5''$ and
derived $N(\rm{CH_3OH}$-$E)=7\times 10^{14}$~cm$^{-2}$ and
$N(\rm{CH_3OH}$-$A)=8\times 10^{14}$~cm$^{-2}$. Our estimate of $n$ in the clumps
agrees with previous studies (e.g. \citealt{2000ApJ...540..886Y}; LS03);
T$_{kin}$ is also similar to previous results for the clump gas (see discussion in LS03)
and to theoretical models \citep{2002ApJ...573..215G}.
The H$_2$ column density ($9\times 10^{22}$~cm$^{-2}$) was 
derived from our observations of the C$^{17}$O(3-2) line, assuming  
$^{17}$O/$^{16}$O$\sim 1790$ \citep{1994ARA&A..32..191W} and $T_{rot}$=70~K. 
The methanol column density then corresponds to a fractional 
abundance 
averaged over the beam of $10^{-9}$, 
typical of dark clouds.
The $3\sigma$ confidence level for $T_{kin}$ ranges between 50 and 75~K, 
and (5--20)$\times 10^6$~cm$^{-3}$ for $n_{\rm H_2}$.

The parameters derived from H$_2$CO differ significantly from the ones
of CH$_3$OH, as high temperature ($T\sim 150$~K) and moderate density
($n\sim 5\times 10^5$~cm$^{-3}$) are needed to reproduce the
data. 
Since all H$_2$CO lines are optically thin, we derived a
beam-averaged column density of $9 \times 10^{13}$~cm$^{-2}$ for
ortho-H$_2$CO and $3 \times 10^{13}$~cm$^{-2}$ for para-H$_2$CO, corresponding to a fractional abundance of $10^{-9}$.
\citet{1996A&A...312..599F} found the same value, although at another position in the Bar.  Our
analysis infers a $T_{kin}$ in agreement with the results of \citet{2003A&A...408..231B} from NH$_3$, but
  higher than the one derived by
\citet{hogerheijde_AaA_294_792_1995}, $85\pm 30$~K, also with formaldehyde, but 
toward a different position. 
The density derived is
substantially higher than that proposed by
\citet{hogerheijde_AaA_294_792_1995} for the interclump medium
($3\times 10^4$~cm$^{-3}$), but it agrees with the value inferred
by \citet{1997A&A...327L...9S} based on CN.  While the $n_{\rm H_2}$ is well-constrained
(1--5$\times 10^5$~cm$^{-3}$), the temperature we derived is a lower limit.

The infrared pumping through an
internal radiation field 
would probably 
affect our results for both molecular species. 
However,
since no model of the IR field is available for the Bar,  no
internal radiation field was taken into account.

Our data suggest that CH$_3$OH is found mainly  in the clumpy
medium, while H$_2$CO traces the interclump material.
Both methanol and formaldehyde 
can form on grain surfaces \citep{2002ApJ...571L.173W}.
It therefore seems remarkable that
they trace different environments. 
 We speculate that in the
interclump gas, methanol, once released from grain surfaces, is
photodissociated to form formaldehyde \citep[one of the possible
photodissociation products, ][] {2000A&AS..146..157L}. Alternatively, 
our observations may indicate a different
formation mechanism for the two species, surface chemistry for CH$_3$OH 
\citep[as no gas-reaction can efficiently form it, ][]{2002luca}, and gas phase chemistry
for H$_2$CO \citep[via the neutral--neutral reaction ${\rm CH}_3+{\rm O}$, e.g.][]{2000A&AS..146..157L}. More detailed
modeling, including release of molecules from grain surfaces in the PDR, is clearly
needed.

\subsection{Sulfur-bearing species}
The excitation analysis of the S-bearing molecules listed in Table~\ref{lineid} 
was performed by fitting  all
the observed transitions with a synthetic spectrum computed under the
LTE assumption, in the manner described by
\citet{comito_ApJS_156_127_2005}.  Since the kinetic temperature, the
source size, and the column density are degenerate parameters in the
optically thin limit, we fixed $T_{kin}$ to the
value derived from H$_2$CO and fit the beam-averaged column density of each
species. For SO$_2$, OCS, and HCS, upper limits to the column densities
were derived, based on the non-detections of lines in our survey range. 
Results are given in Table~\ref{s-mol}. The uncertainties on the column densities are derived with a
$\chi^2$ analysis and correspond 
to the $3\sigma$ confidence level. CS and its isotopologs are not included in our analysis,
because of the anomalous $^{33}$S/$^{34}$S ratio of $\sim3$ we observed \citep[for comparison, see ][]{1996A&A...305..960C}. Observations of other rotational transitions are needed
to constrain this ratio reliably.

\citet[][ hereafter SD95]{1995ApJS...99..565S}  studied the chemistry of PDRs produced in a dense molecular 
cloud exposed to intense far-UV radiation fields. The density they employ ($n_H=10^6$~cm$^{-3}$) is higher than
the one relevant here for the interclump gas; discrepancies in the results may be due to this. Their chemical network does not include NS and H$_2$CS, 
but all the other S-bearing molecules observed here are found there. 

According to SD95, the SO$^+$/SO ratio (observed 0.2 from
Table~\ref{s-mol}) can be used as diagnostic of the different chemical
zones in the cloud and changes between 0.23 at $A_v=1.5$ and 0.0028 at
$A_v=3$. 
Assuming that the Bar is edge-on, based on the distance between the
ionization front and our position ($\sim 30''$), and assuming an H$_2$
density of $10^5$~cm$^{-3}$, we derived a visual extinction of $\ge
10$~mag 
\citep{1982ApJ...262..590F}.  Given the complex geometry of the Orion
Bar, which has edge-on (but oblique) and face-on parts
\citep{hogerheijde_AaA_294_792_1995}, it is likely that together with
the molecular material at this extinction, a hotter, more diffuse layer
of gas, corresponding to a face-on surface layer, contributes to the
emission we observe.  SD95 predict SO$_2$ to be
more abundant than SO and SO$^+$ at high $A_v$, while our upper limit
indicates that SO$_2$ is less abundant.  Similar to the result found
for SO$^+$/SO, our estimates of SO$_2$/SO and SO$_2$/SO$^+$ suggest
that these species have a significant contribution from a hotter layer of gas with lower
extinction than the molecular layer.

\citet{fuente_AaA_406_899_2003} report different column densities for SO$^+$ and
SO$_2$; however, their SO$^+$/SO$_2$ ratio ranges
between 1.2 at the IR front and 0.4 at (20$''$,-20$''$) from it, which is
similar to
our value of 0.3. 
 \citet{jansen_AaA_303_541_1995} found values of N(H$_2$S) between $8.9\times 10^{12}$~cm$^{-2}$ and
$3.9\times 10^{14}$~cm$^{-2}$ across the Bar, with our value of $1.3\times 10^{13}$~cm$^{-2}$ lying in between.
Discrepancies in the absolute values of the column densities may depend on the source sizes, on
temperatures, and on the total H$_2$ column density at the observed position.

\begin{table}[h!]
\caption{Column densities of S-bearing molecules}\label{s-mol}
\begin{tabular}{lclr }
\hline\hline
Species&$N$ [cm$^{-2}$]&Species&$N$ [cm$^{-2}$]\\
NS&$(1.3\pm 0.3) 10^{13}$    &SO$_2$&$ <3~10^{13}$\\
SO&$(1.0\pm 0.2) 10^{14}$    &HCS&$<4~10^{14}$   \\
SO$^+$&$(2.0\pm 0.5) 10^{13}$&OCS&$<4~10^{13}$   \\
H$_2$S&$(2.5\pm 1.0) 10^{13}$\\
H$_2$CS&$(2.5\pm 0.4) 10^{13}$\\
HCS$^+$&$(2.5\pm 0.7) 10^{12}$\\

\hline\hline
\end{tabular}
\end{table}
\section{Detection of a deuterated molecule : DCN}

\begin{figure}
\begin{center}
\includegraphics[width=4cm,angle=-90]{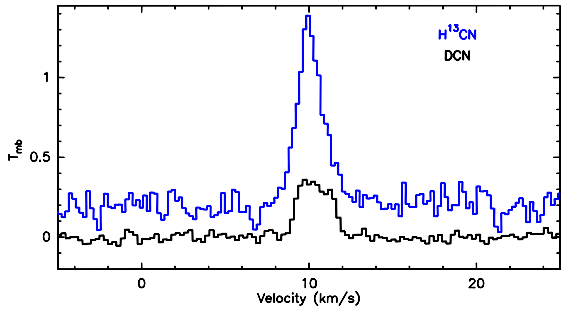}
\label{dcn}
\caption{H$^{13}$CN(4-3) and DCN(4-3) lines. For clarity, the H$^{13}$CN line has been translated by 0.2 along the $y$-axis.}
\end{center}
\end{figure}

Molecular deuteration studies have been extremely powerful
in probing the physical history of sources. The deuterium isotopic
ratio in molecules can indeed be enhanced in low-temperature
environments compared to the cosmic D/H ratio
\citep[$\sim$\,10$^{-5}$,][]{Linsky03}. This enhancement proceeds
initially from the transfer of deuterium from the main reservoir HD to
two reactive ions (H$_2$D$^+$ and CH$_2$D$^+$) through exothermic
reactions with H$_3^+$ and CH$_3$$^+$ \citep[][ and references therein]{2000A&A...361..388R}. The endothermicity of the
backward reactions (respectively by 220\,K and 370\,K) ensures that
the H$_2$D$^+$/H$_3^+$ and CH$_2$D$^+$/CH$_3^+$ ratios are
significantly enhanced at low temperatures.  Deuterium is then
channeled to other molecules by gas-phase reactions with these two
ions and by grain surface reactions. Highly-deuterated molecules can
also be found in hot gas, where they are then out-of-equilibrium
fossils of an earlier cold phase \citep[e.g. in hot corinos around
low-mass protostars, where they are believed to be remnants of the
prestellar cold and depleted chemistry;][]{Parise04,Parise06}.

The DCN(4-3) transition at 289.645\,GHz (E$_{up}$\,=\,34.56\,K) was detected in our survey, and found to be
very bright (Fig.~\ref{dcn}).
This detection, together with the
detection of DCO$^+$ in the Horsehead nebula (Pety et al. 2006, in preparation), is
opening up the study of fractionation processes in PDRs. The
H$^{13}$CN(4-3) transition (E$_{up}$\,=\,41.76\,K) was also
detected at 345.340\,GHz. 
The H$^{13}$CN emission was shown from
interferometric observations (LS03) to be
mostly associated with clumps in the PDR. 
The DCN emission is thus very
likely to also be associated with clumps, as DCN would not survive the
high temperatures of the interclump gas. The linewidths of DCN(4-3) and H$^{13}$CN(4-3) are comparable and non-Gaussian
profiles are detected in both cases.
We thus assume that
H$^{13}$CN and DCN trace the same region. As the energies of the two
lines are very similar, the derived abundance ratio is nearly
independent of the temperature. Assuming LTE and a $^{12}$C/$^{13}$C
ratio of 70, we derived  beam-averaged DCN and H$^{13}$CN column densities
between $8.4\times 10^{11}$ and $1.7\times 10^{12}$~cm$^{-2}$, and  between $1.7\times 10^{12}$ and $2.8\times 10^{12}$~cm$^{-2}$, respectively, 
in the range of temperatures 30--150~K. This corresponds to a DCN/HCN ratio of
0.7-0.9\%, which is 
an intermediate value between that
observed in warm gas \citep[Orion Hot Core, 0.1\%,][]{1992A&A...256..595S}, on
the one hand, and in dark clouds \citep[L134, 5\%,][]{Turner01} or cold
(30-50\,K) gas of the OMC1 ridge region \citep[1-6\%][]{1992A&A...256..595S}, on
the other.

In hot cores/corinos, where the gas has not yet had
time to return to equilibrium (which would require timescales of
$\sim$ 10$^4$ yrs), abundant deuterated molecules can be
considered as remnants of earlier cold prestellar chemistry, most
probably stored in the ices on grain surfaces and then released when
the protostar heats its surroundings. In contrast, the gas in the shielded clumps of the bar might be
in steady state. In this case, the observed deuterated molecules 
have to be present-day product molecules. The detection of DCN in this
relatively warm environment, where H$_2$D$^+$ would not survive, is in
this sense quite striking and is a strong hint of the efficiency of
the fractionation processes occurring via the CH$_2$D$^+$ channel,
which survives higher temperatures than H$_2$D$^+$, due to the higher
exothermicity of the CH$_3^+$ reaction with HD \citep{Turner01}. The
Orion Bar thus  appears  a unique reference to test the fractionation
reactions involving CH$_2$D$^+$. Knowing the spatial distribution of
the DCN/HCN isotopic ratio would then be an interesting tool for
understanding the effect of extinction on the fractionation processes.

\begin{acknowledgements}
BP is grateful to the Alexander von Humboldt Foundation for a  Research
fellowship 
\end{acknowledgements}

\bibliographystyle{aa}
\bibliography{5555}

\begin{thebibliography}{32}
\expandafter\ifx\csname natexlab\endcsname\relax\def\natexlab#1{#1}\fi

\bibitem[{{Batrla} \& {Wilson}(2003)}]{2003A&A...408..231B}
{Batrla}, W. \& {Wilson}, T.~L. 2003, \aap, 408, 231

\bibitem[{{Chin} {et~al.}(1996){Chin}, {Henkel}, {Whiteoak}, {Langer}, \&
  {Churchwell}}]{1996A&A...305..960C}
{Chin}, Y.-N., {Henkel}, C., {Whiteoak}, J.~B., {Langer}, N., \& {Churchwell},
  E.~B. 1996, \aap, 305, 960

\bibitem[{{Comito} {et~al.}(2005){Comito}, {Schilke}, {Phillips}, {Lis},
  {Motte}, \& {Mehringer}}]{comito_ApJS_156_127_2005}
{Comito}, C., {Schilke}, P., {Phillips}, T.~G., {et~al.} 2005, \apj Suppl. S.,
  156, 127

\bibitem[{{Frerking} {et~al.}(1982){Frerking}, {Langer}, \&
  {Wilson}}]{1982ApJ...262..590F}
{Frerking}, M.~A., {Langer}, W.~D., \& {Wilson}, R.~W. 1982, \apj, 262, 590

\bibitem[{{Fuente} {et~al.}(2003){Fuente}, {Rodr{\'i}guez-Franco},
  {Garc{\'i}a-Burillo}, {Mart{\'i}n-Pintado}, \&
  {Black}}]{fuente_AaA_406_899_2003}
{Fuente}, A., {Rodr{\'i}guez-Franco}, A., {Garc{\'i}a-Burillo}, S.,
  {Mart{\'i}n-Pintado}, J., \& {Black}, J.~H. 2003, \aap, 406, 899

\bibitem[{{Fuente} {et~al.}(1996){Fuente}, {Rodriguez-Franco}, \&
  {Martin-Pintado}}]{1996A&A...312..599F}
{Fuente}, A., {Rodriguez-Franco}, A., \& {Martin-Pintado}, J. 1996, \aap, 312,
  599

\bibitem[{{Gorti} \& {Hollenbach}(2002)}]{2002ApJ...573..215G}
{Gorti}, U. \& {Hollenbach}, D. 2002, \apj, 573, 215

\bibitem[{{Green}(1991)}]{1991ApJS...76..979G}
{Green}, S. 1991, \apjs, 76, 979

\bibitem[{{Hogerheijde} {et~al.}(1995){Hogerheijde}, {Jansen}, \& {van
  Dishoeck}}]{hogerheijde_AaA_294_792_1995}
{Hogerheijde}, M.~R., {Jansen}, D.~J., \& {van Dishoeck}, E.~F. 1995, \aap,
  294, 792

\bibitem[{{Hollenbach} \& {Tielens}(1997)}]{hollenbach_ARA&A_35_179_1997}
{Hollenbach}, D.~J. \& {Tielens}, A.~G.~G.~M. 1997, \araa, 35, 179

\bibitem[{{Jansen} {et~al.}(1995){Jansen}, {Spaans}, {Hogerheijde}, \& {van
  Dishoeck}}]{jansen_AaA_303_541_1995}
{Jansen}, D.~J., {Spaans}, M., {Hogerheijde}, M.~R., \& {van Dishoeck}, E.~F.
  1995, \aap, 303, 541

\bibitem[{{Le Teuff} {et~al.}(2000){Le Teuff}, {Millar}, \&
  {Markwick}}]{2000A&AS..146..157L}
{Le Teuff}, Y.~H., {Millar}, T.~J., \& {Markwick}, A.~J. 2000, \aaps, 146, 157

\bibitem[{{Leurini} {et~al.}(2004){Leurini}, {Schilke}, {Menten}, {Flower},
  {Pottage}, \& {Xu}}]{leurini_A&A_422_573_2004}
{Leurini}, S., {Schilke}, P., {Menten}, K.~M., {et~al.} 2004, \aap, 422, 573

\bibitem[{{Linsky}(2003)}]{Linsky03}
{Linsky}, J.~L. 2003, Space Science Reviews, 106, 49

\bibitem[{{Lis} \& {Schilke}(2003)}]{lis_ApJL_597_L145_2003}
{Lis}, D.~C. \& {Schilke}, P. 2003, \apj, 597, L145, LS03

\bibitem[{{Luca} {et~al.}(2002){Luca}, {Voulot}, \& {Gerlich}}]{2002luca}
{Luca}, A., {Voulot}, D., \& {Gerlich}, D. 2002, in WDS'02 Proceedings of
  Contributed Papers, Part II, Safrankova (ed), Matfyzpress, 294

\bibitem[{{M{\" u}ller} {et~al.}(2001){M{\" u}ller}, {Thorwirth}, {Roth}, \&
  {Winnewisser}}]{mueller_cdms}
{M{\" u}ller}, H.~S.~P., {Thorwirth}, S., {Roth}, D.~A., \& {Winnewisser}, G.
  2001, \aap, 370, L49

\bibitem[{{Parise} {et~al.}(2004){Parise}, {Castets}, {Herbst}, {Caux},
  {Ceccarelli}, {Mukhopadhyay}, \& {Tielens}}]{Parise04}
{Parise}, B., {Castets}, A., {Herbst}, E., {et~al.} 2004, \aap, 416, 159

\bibitem[{{Parise} {et~al.}(2006){Parise}, {Ceccarelli}, {Tielens}, {Castets},
  {Cazaux}, {Lefloch}, \& {Maret}}]{Parise06}
{Parise}, B., {Ceccarelli}, C., {Tielens}, A., {et~al.} 2006, \aap, in press

\bibitem[{{Pety} {et~al.}(2005){Pety}, {Teyssier}, {Foss{\'e}}, {Gerin},
  {Roueff}, {Abergel}, {Habart}, \& {Cernicharo}}]{pety_AaA_435_885_2005}
{Pety}, J., {Teyssier}, D., {Foss{\'e}}, D., {et~al.} 2005, \aap, 435, 885

\bibitem[{{Pickett} {et~al.}(1998){Pickett}, {Poynter}, {Cohen}, {Delitsky},
  {Pearson}, \& {Muller}}]{pickett_JMolSpectrosc_60_883_1998}
{Pickett}, H.~M., {Poynter}, R.~L., {Cohen}, E.~A., {et~al.} 1998, J. Quant.
  Spectrosc. Radiat. Transfer, 60

\bibitem[{{Pottage} {et~al.}(2002){Pottage}, {Flower}, \&
  {Davis}}]{2002JPhB...35.2541P}
{Pottage}, J.~T., {Flower}, D.~R., \& {Davis}, S.~L. 2002, J. Phys. B: At. Mol.
  Phys., 35, 2541

\bibitem[{{Roberts} \& {Millar}(2000)}]{2000A&A...361..388R}
{Roberts}, H. \& {Millar}, T.~J. 2000, \aap, 361, 388

\bibitem[{{Schilke} {et~al.}(2001){Schilke}, {Pineau des For{\^ e}ts},
  {Walmsley}, \& {Mart{\'{\i}}n-Pintado}}]{schilke_A&A_372_291_2001}
{Schilke}, P., {Pineau des For{\^ e}ts}, G., {Walmsley}, C.~M., \&
  {Mart{\'{\i}}n-Pintado}, J. 2001, \aap, 372, 291

\bibitem[{{Schilke} {et~al.}(1992){Schilke}, {Walmsley}, {Pineau Des Forets},
  {Roueff}, {Flower}, \& {Guilloteau}}]{1992A&A...256..595S}
{Schilke}, P., {Walmsley}, C.~M., {Pineau Des Forets}, G., {et~al.} 1992, \aap,
  256, 595

\bibitem[{{Simon} {et~al.}(1997){Simon}, {Stutzki}, {Sternberg}, \&
  {Winnewisser}}]{1997A&A...327L...9S}
{Simon}, R., {Stutzki}, J., {Sternberg}, A., \& {Winnewisser}, G. 1997, \aap,
  327, L9

\bibitem[{{Sternberg} \& {Dalgarno}(1995)}]{1995ApJS...99..565S}
{Sternberg}, A. \& {Dalgarno}, A. 1995, \apjs, 99, 565, SD95

\bibitem[{{St\"orzer} {et~al.}(1995){St\"orzer}, {Stutzki}, \&
  {Sternberg}}]{stoerzer_A&A_296_9_1995}
{St\"orzer}, H., {Stutzki}, J., \& {Sternberg}, A. 1995, \aap, 296, L9

\bibitem[{{Turner}(2001)}]{Turner01}
{Turner}, B.~E. 2001, \apjs, 136, 579

\bibitem[{{Watanabe} \& {Kouchi}(2002)}]{2002ApJ...571L.173W}
{Watanabe}, N. \& {Kouchi}, A. 2002, \apjl, 571, L173

\bibitem[{{Wilson} \& {Rood}(1994)}]{1994ARA&A..32..191W}
{Wilson}, T.~L. \& {Rood}, R. 1994, \araa, 32, 191

\bibitem[{{Young Owl} {et~al.}(2000){Young Owl}, {Meixner}, {Wolfire},
  {Tielens}, \& {Tauber}}]{2000ApJ...540..886Y}
{Young Owl}, R.~C., {Meixner}, M.~M., {Wolfire}, M., {Tielens}, A.~G.~G.~M., \&
  {Tauber}, J. 2000, \apj, 540, 886

\end{thebibliography}

\end{document}